\documentclass[lettersize,journal]{IEEEtran}
\usepackage{amssymb, amsfonts}
\usepackage[fleqn]{amsmath}
\usepackage{algorithmic}
\usepackage{array}
\usepackage{textcomp}
\usepackage{cite}
\usepackage{stfloats}
\usepackage{url}
\usepackage{verbatim}
\usepackage{graphicx}
\usepackage{booktabs}
\usepackage{makecell}
\def\BibTeX{{\rm B\kern-.05em{\sc i\kern-.025em b}\kern-.08em
    T\kern-.1667em\lower.7ex\hbox{E}\kern-.125emX}}
\usepackage{balance}
\IEEEoverridecommandlockouts
\usepackage[switch]{lineno}

\usepackage{xcolor}
\usepackage{subfig}
\usepackage{float}

\begin{document}
\title{Predicting Trust Dynamics Type Using Seven Personal Characteristics}
\author{Hyesun Chung, X. Jessie Yang
\thanks{This work has been submitted to the IEEE for possible publication. Copyright may be transferred without notice, after which this version may no longer be accessible. (Corresponding author: X. Jessie Yang).}}



\maketitle

\begin{abstract}
This study aims to explore the associations between individuals' trust dynamics in automated/autonomous technologies and their personal characteristics, and to further examine whether personal characteristics can be used to predict a user's trust dynamics type.
We conducted a human-subject experiment (N=130) in which participants performed a simulated surveillance task assisted by an automated threat detector. Using a pre-experimental survey covering 12 constructs and 28 dimensions, we collected data on participants' personal characteristics. Based on the experimental data, we performed k-means clustering and identified three trust dynamics types. Subsequently, we conducted one-way Analyses of Variance to evaluate differences among the three trust dynamics types in terms of personal characteristics, behaviors, performance, and post-experimental ratings. Participants were clustered into three groups, namely Bayesian decision makers, disbelievers, and oscillators. Results showed that the clusters differ significantly in seven personal characteristics: masculinity, positive affect, extraversion, neuroticism, intellect, performance expectancy, and high expectations. The disbelievers tend to have high \textit{neuroticism} and low \textit{performance expectancy}. The oscillators tend to have higher scores in \textit{masculinity}, \textit{positive affect}, \textit{extraversion}, and \textit{intellect}.
We also found significant differences in behaviors, performance, and post-experimental ratings across the three groups. The disbelievers are the least likely to blindly follow the recommendations made by the automated threat detector. Based on the significant personal characteristics, we developed a decision tree model to predict the trust dynamics type with an accuracy of 70\%. This model offers promising implications for identifying individuals whose trust dynamics may deviate from a Bayesian pattern.

\end{abstract}


\section{Introduction}
\IEEEPARstart{T}{rust} in automated/autonomous technologies\footnote{In this paper, we use the automated/autonomous technologies notation while acknowledging the difference between automation and autonomy \cite{o2022human, endsley2017here}} has received growing attention due to the increasing adoption of them. In early work on trust, an extensive body of research has focused on identifying factors that influence trust
\cite{hoff2015trust, hancock2021evolving, kaplan2023trust, schaefer2016meta, hancock2011meta}. In these works, a person's trust in automated/autonomous technologies is typically measured at specific time points, i.e., \textit{snapshot} trust, predominantly at the end of an experiment after a series of interactions and at the beginning of an experiment.

More recently, acknowledging that a person's trust can change dynamically while interacting with automated/autonomous technologies, there is a shift of research focus from snapshot trust to trust dynamics -- how humans' trust in automated/autonomous technologies forms and evolves due to moment-to-moment interaction with it \cite{DeVisser_IJSR, yang2023toward,yang_trust_2023}. 
Real-time trust prediction models \cite{bhat_evaluating_2023, guo2021modeling, Bhat2022_RAL, Chen:2018ug, McMahon2020_IFAC, Liu2021_ITSC, guo_enabling_2023} are developed and distinct types of trust dynamics are revealed \cite{guo2021modeling, McMahon2020_IFAC, Bhat2022_RAL, Liu2021_ITSC}. 
McMahon et al. \cite{McMahon2020_IFAC} identified two distinct types of trust dynamics based on trust-dependent behavior: followers and preservers. 
In another study, Liu et al. \cite{Liu2021_ITSC} revealed two groups of trust dynamics: confident and skeptical. Compared to the skeptical group, the confident group exhibited higher initial trust, slower trust erosion following trust violations, and higher trust recovery. Guo and Yang \cite{guo2021modeling} found three types of trust dynamics in an \emph{episodic} task setting: Bayesian decision maker, oscillator, and disbeliever. In a follow-up study, the same three types were revealed in a \emph{sequential} task setting \cite{Bhat2022_RAL}. The Bayesian decision makers update their trust in a Bayesian manner, the oscillators' trust fluctuates dramatically, and the disbelievers display extremely low trust in automation. 

Despite existing research efforts on trust dynamics, several research gaps remain. First, prior studies on clustering trust dynamics have predominantly focused on automation with reliability levels above 70\% \cite{guo2021modeling, Bhat2022_RAL, McMahon2020_IFAC}. It remains unclear whether the three types of trust dynamics would still be consistently observed if automation reliability falls below 70\%. Second, while research on snapshot trust has revealed significant associations between personal characteristics and post-experimental trust \cite{zhou2020effects, sharan2020effects, huang2020if}, as well as between personal characteristics and trust propensity (i.e., an individual's inherent tendency to trust automated/autonomous technologies) \cite{tang2022users, huang2018users, chien2016relation}, the relationships between trust dynamics and personal characteristics have not been fully explored. This gap limits our understanding of which personal traits influence trust dynamics the most and how they do so. Third, building on the second gap, the potential of leveraging individuals' personal traits to predict their trust dynamics remains largely unexplored.

To address the aforementioned gaps, 
this study seeks to answer the following three key research questions:

\begin{enumerate}
    \item What types of trust dynamics can be identified when automation reliability falls below 70\%?
    \item Are there significant differences in personal characteristics across different types of trust dynamics?
    \item Can key personal characteristics predict the type of trust dynamics a user will exhibit?
    
\end{enumerate}

We first reviewed key literature review papers on trust in automated/autonomous technologies and synthesized factors that affect \textit{snapshot} trust \cite{hoff2015trust, hancock2021evolving, kaplan2023trust, hancock2011meta, schaefer2016meta}. This process resulted in the development of a list comprising 12 personal characteristics constructs encompassing 28 dimensions. The survey to measure these was administered before the experiment. During the experiment, participants' dynamic trust data was recorded. Using the experimental data, we conducted k-means clustering and identified three clusters of trust dynamics: Bayesian decision maker, disbeliever, and oscillator. Subsequently, we performed multiple one-way Analyses of Variance (ANOVAs) to examine differences among the three clusters. The clusters differed significantly in seven personal characteristics (masculinity, positive affect, extraversion, neuroticism, intellect, performance expectancy, high expectations), behaviors, performance, and post-experimental ratings. Lastly, we developed a decision tree model to predict trust dynamics type based on the seven key personal characteristics. The decision tree achieved an overall accuracy of 70.0\%.

\section{Related Work}

This section reviews three research areas that motivated our study: the shift from snapshot trust to trust dynamics, the link between personal traits and snapshot trust, and studies on clustering trust dynamics.

\subsection{From trust to trust dynamics}
Trust in automated/autonomous technologies has been receiving growing research attention \cite{chung2024developing}. Early works on this topic predominantly embraced a static perspective on trust, gauging it through pre- or post-experimental surveys. Numerous factors have been identified as antecedents of post-experimental trust. See \cite{hoff2015trust, hancock2021evolving, kaplan2023trust, schaefer2016meta, hancock2011meta, wischnewski2023measuring} for a full list of factors.

More recently, acknowledging that a person's trust can change dynamically while interacting with automated/autonomous technologies, there is a shift of research focus from snapshot trust to trust dynamics. A growing body of research studies is focused on understanding how humans' trust in automated/autonomous technologies forms and evolves due to moment-to-moment interaction with it \cite{DeVisser_IJSR, yang2023toward, yang_trust_2023, wischnewski2023measuring}.

One line of research on trust dynamics explores how trust is developed, violated, and recovered during the course of interaction \cite{lee1994trust, manzey2012human, yang2023toward, DeVisser_IJSR, esterwood2023three, desai2013impact}. Trust is established and developed as individuals learn from the outcomes of their interactions, whereas trust violations occur following negative experiences \cite{esterwood2023three, esterwood2022literature}. In cases where trust violations occur, efforts can be made to restore trust. These efforts often involve various actions/strategies such as apologies \cite{mahmood2022owning}, expressing regrets \cite{kox2021trust}, denials \cite{kohn2019consequences}, explanations \cite{natarajan2020effects, kox2021trust}, and promises \cite{albayram2020investigating}.

The second line of research is focused on developing real-time trust prediction models \cite{xu2015optimo, guo2021modeling, Bhat2022_RAL, Chen:2018ug, azevedo2021real}. Examples include the online probabilistic trust inference model (OPTIMo) by \cite{xu2015optimo}, the Beta distribution model by \cite{guo2021modeling}, and the Bayesian model combining Gaussian processes and recurrent neural networks by \cite{soh2020multi}. For a detailed review, please refer to \cite{kok2020trust}.

\subsection{Associations between personal characteristics and post-experimental (snapshot) trust}

Extensive research has investigated personal factors that influence (snapshot) trust in automated/autonomous technologies, predominantly measured at the end of an experiment after a series of interactions. Results show that human-related factors are significantly related to post-experimental trust \cite{hancock2021evolving, kaplan2023trust, lin2022changes}, which can be classified into characteristics-based and ability-based variables \cite{hancock2021evolving}. 

The characteristics-based factors relate to individual traits, such as personality, culture, mood, attitudes toward automation, propensity to trust, risk propensity, and decision-making style. Personality traits like extraversion are known to positively correlate with trust \cite{Merritt:2008ds, zhou2020effects}, while neuroticism shows a negative relationship \cite{zhou2020effects, sharan2020effects}. In terms of cultural characteristics, previous studies proposed that individuals who hold stronger beliefs in individualism and horizontal collectivism—which emphasizes equality, uniqueness, and self-reliance—tend to have higher trust in automated/autonomous technologies, in contrast to those who value vertical collectivism, which emphasizes authority and individual self-sacrifice \cite{huang2018users, chien2018effect}. Positive moods \cite{stokes2010accounting}, positive attitudes toward automation, and high trust propensity \cite{merritt2013trust, bhat_evaluating_2023} are also known to correlate with higher trust ratings. Risk propensity, an individual's tendency to take risks \cite{meertens2008measuring}, and decision-making style also influence trust \cite{stuck2021role, mcbride2012impact}.

The ability-based variables encompass factors associated with a user's skills and competence, including attentional capacity and self-confidence \cite{hancock2021evolving}. 
Empirical studies showed a negative relationship between attentional capacity and trust \cite{chen2009effects}. 
Likewise, self-confidence is known to have a negative association. People with relatively higher self-confidence were likely to have lower trust and thus prefer to exert full control over a given task rather than following the automated/autonomous technologies \cite{lee1994trust, de2003effects}.

In summary, numerous human-related variables have significant associations with post-experimental trust. The insights from previous research have led us to examine the associations between personal characteristics and trust dynamics, and validate whether similar relationships exist.

\subsection{Clusters of trust dynamics}
 
Notably, recent studies have revealed the existence of different types of trust dynamics \cite{guo2021modeling, McMahon2020_IFAC, Bhat2022_RAL, Liu2021_ITSC}. McMahon et al. \cite{McMahon2020_IFAC} identified two distinct clusters, followers and preservers, based on participants' trust-dependent behaviors when interacting with a recommendation robot with a reliability of 70\%. Followers were characterized by a high propensity to trust and a greater likelihood of complying with recommendations. In contrast, preservers displayed lower levels of trust and were less likely to follow recommendations. Liu et al. \cite{Liu2021_ITSC} categorized participants into skeptical and confident groups based on variables derived from 12 different features in a takeover situation. Compared to the skeptical group, the confident group exhibited higher initial trust, slower trust erosion following trust violations, and greater trust recovery. In this study, automation reliability was defined by how aggressively or smoothly the vehicle decelerated as it approached the stop line. Guo and Yang \cite{guo2021modeling} identified three types of trust dynamics in an \emph{episodic} task setting where automation reliability ranged from 70 to 90\%: Bayesian decision maker, oscillator, and disbeliever. The Bayesian decision makers update their trust in a Bayesian manner, the oscillators' trust fluctuates dramatically, and the disbelievers display very low trust in automation. Bhat et al. \cite{Bhat2022_RAL} found the same three types in a \emph{sequential} task setting where the reliability was 85\%. This body of clustering research underscores the suitability of unsupervised clustering methods for revealing unique trust dynamics types.

It is important to note that the majority of the clustering studies have focused on automation reliability levels above 70\%. Exploring trust dynamics when automation falls below this threshold would provide valuable insights into user trust dynamics with imperfect automation.

In addition, there are studies that incorporated behavior-related variables (e.g., compliance in and reliance on automation) into clustering
\cite{McMahon2020_IFAC, Liu2021_ITSC}. While it might seem reasonable to include these metrics to classify individuals' reliance behaviors, it is essential to differentiate between trust as an attitude and trusting behaviors \cite{lee2004trust}. Considering trust as a behavior could introduce potential confounders, such as workload, situational awareness, and operator self-confidence \cite{lee2004trust, lee1994trust}. Therefore, our study placed a primary focus on trust attitudes. We aimed to identify trust dynamics clusters based solely on users' trust attitudes and investigate whether these distinct trust dynamics clusters exhibit unique patterns of reliance and compliance behaviors.

\subsection{Associations between personal characteristics and \emph{trust dynamics}}

Despite growing research on clustering trust dynamics, limited attention has been directed toward exploring the associations between personal characteristics and trust dynamics. One study \cite{Bhat2022_RAL} attempted this but only covered a very limited subset of personal characteristics, including personality, perfect automation schema, and propensity to trust. In comparison to the extensive body of empirical research evaluating the impacts of various personal factors on snapshot trust, a significant research gap is apparent. Specifically, how trust changes occur and the extent to which they are dynamic and unpredictable may be associated with personal characteristics. However, these relationships remain unresolved due to a lack of empirical evidence. To address this, our study compiled a list of personal characteristics previously shown to affect snapshot trust and, therefore, potentially influence an individual's trust dynamics.

\section{METHOD}
In this study, we conducted a human-subject experiment with 130 participants.
The study adhered to the American Psychological Association's code of ethics and was approved by the institutional review board.

\subsection{Participants}
We collected data from 130 participants (average age=22.6 years, Standard Deviation=3.5 years), all of whom had normal or corrected-to-normal vision and hearing. 
Participants received a base compensation of \$20 upon completing the experiment, with an additional performance-based bonus ranging from \$2.50 to \$10.

\subsection{Apparatus and stimuli}

In the experiment, participants collaborated with a swarm of drones to complete a surveillance task at $100$ sites (i.e. 100 trials), with each trial lasting 10 seconds. During each trial, participants simultaneously performed two tasks: a compensatory tracking task and a detection task, identifying a potential threat from the photo feeds captured by the drones. At any given time, they could only view either the tracking task or the detection task display and had to toggle between the two.

\begin{figure}[!ht]
    \centering\subfloat[``Danger'' alert]{\includegraphics[scale = 0.16]{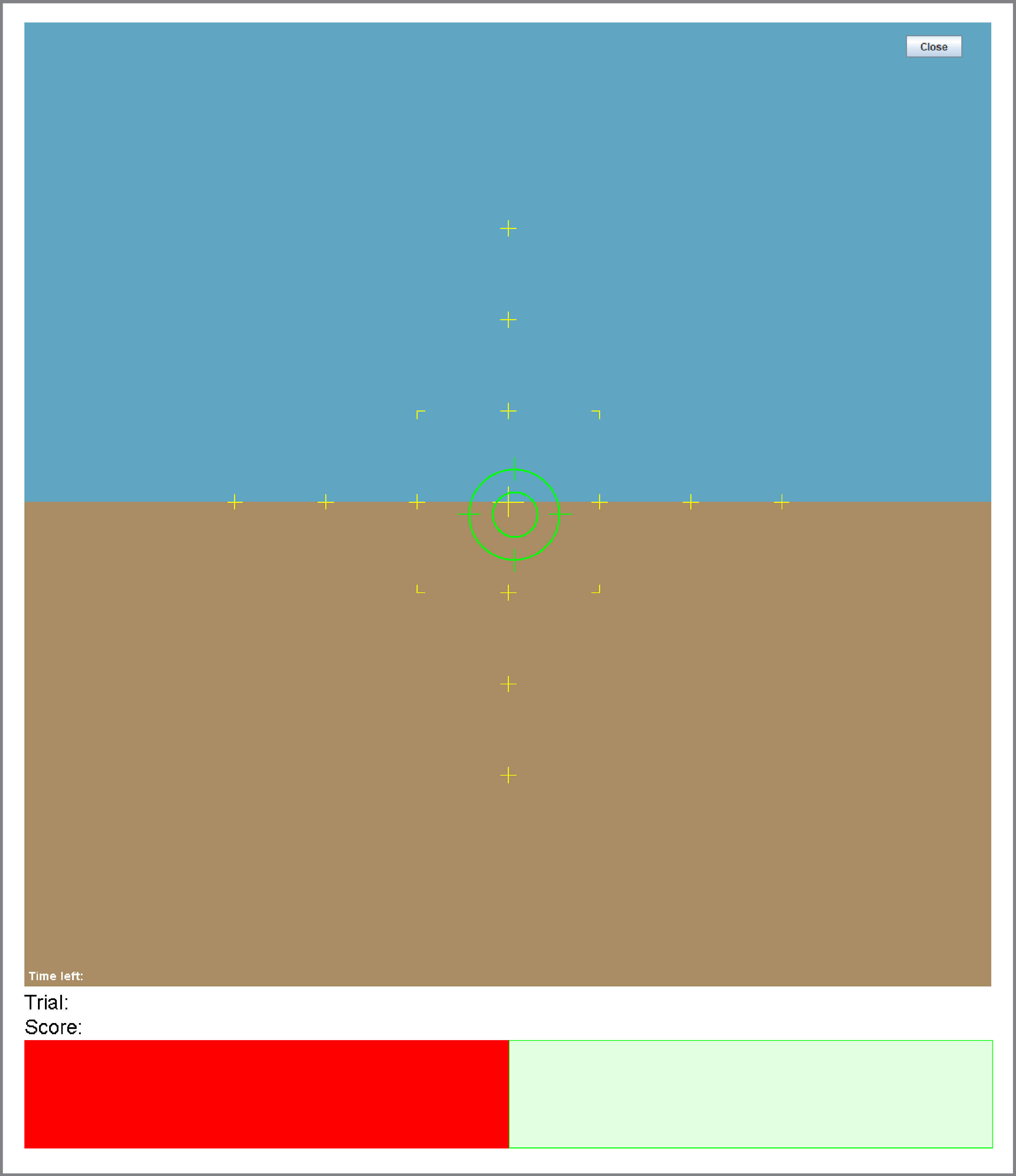}}
    \hspace{0.01 em}
    \subfloat[``Clear'' alert]{\includegraphics[scale = 0.16]{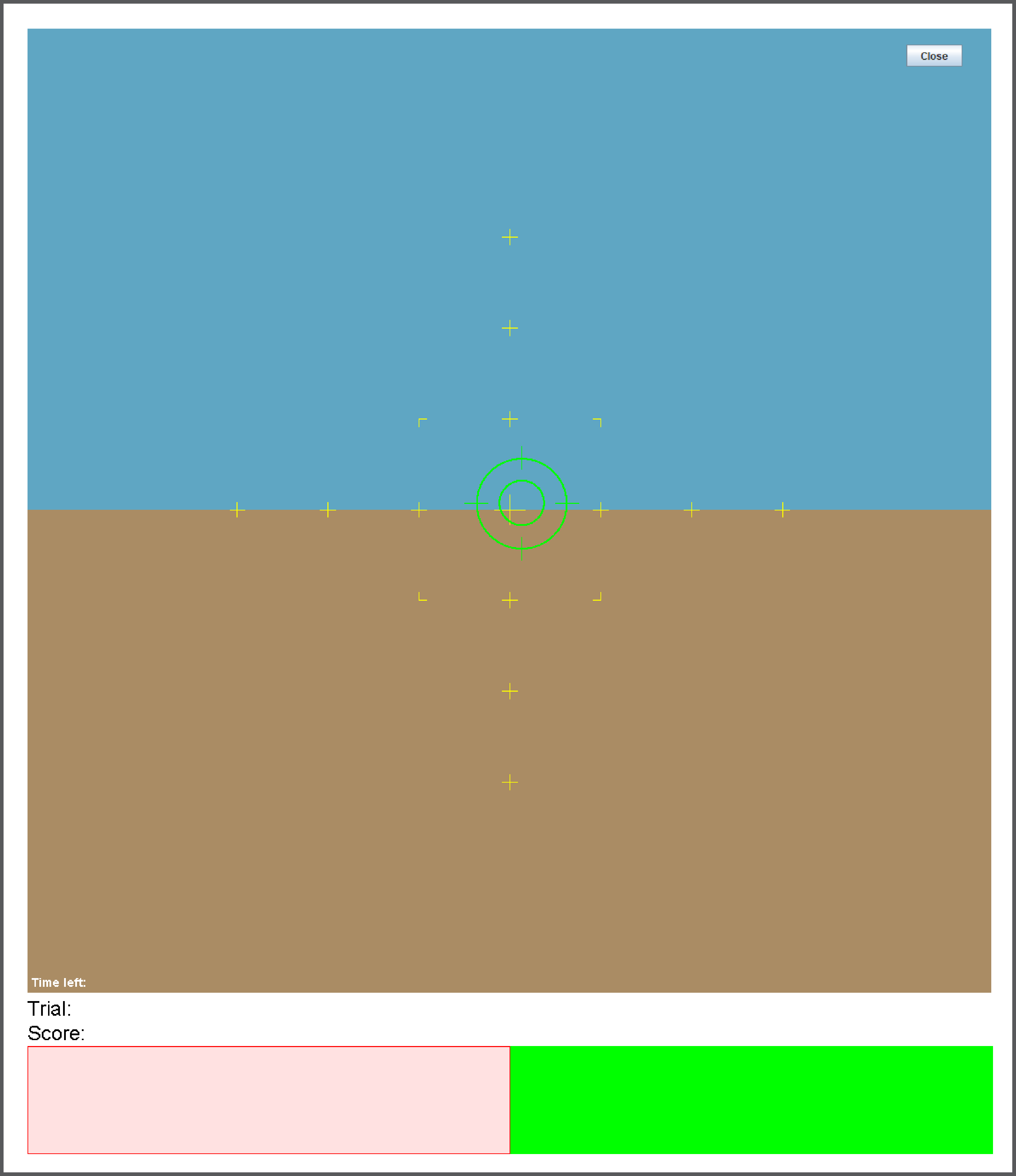}}
    \caption{Tracking task display: In each trial, an automated threat detector provides a visual alert on the bottom bar of the display, indicating whether it identifies a threat (a) or no threat (b).}
    \label{fig:tracking}
\end{figure}

\textbf{\textit{Tracking task}}. 
The tracking task was constructed based on the PEBL (Psychology Experiment Building Language) compensatory tracker task. Participants used a joystick to control the movement of a green circle that drifted randomly on the screen. Their objective was to keep the circle toward a crosshair located at the center of the display (see Fig. \ref{fig:tracking}). 

\begin{figure}[h]
  \centering
  \includegraphics[scale=0.38]{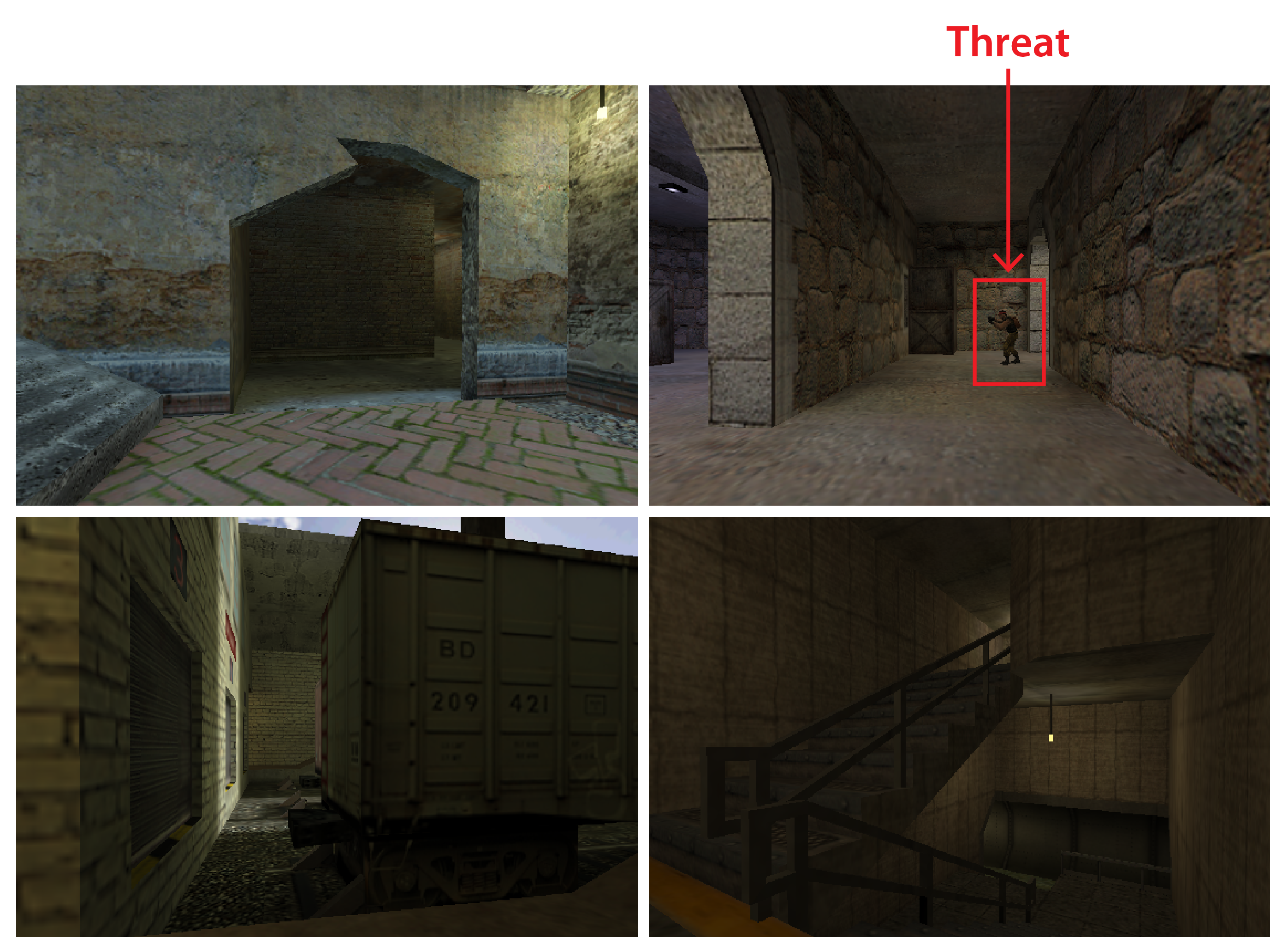}
  \caption{Example of a set of four images presented to one participant at one trial. A threat is located in the top right image. Note that the red box is for illustration purposes. No red box was presented in the real experiment.}
  \label{fig:detection}
\end{figure}

\textbf{\textit{Detection task}}. 
For every trial, participants received a new set of four static images from the drones for threat detection (see Fig. \ref{fig:detection}). They were instructed to acknowledge the presence of a threat by pressing the ``Report'' button on the joystick. No action was required if no threat was present.

The threat being detected was represented by a human figure. In the experiment, if a threat was present, it was indicated by a single person appearing in one of the four images. Multiple threats (i.e., multiple persons) appearing in one image or across multiple images were prohibited.


Participants were assisted by an imperfect automated threat detector for the detection task. 
If the detector identified a threat, a visual alert in the form of a red bar appeared on the tracking display (Fig. \ref{fig:tracking}(a)) at the start of a new trial. Simultaneously, an auditory alert was triggered, consisting of the recorded warning ``Danger,'' played through an external speaker. Conversely, if the detector did not identify a threat, a visual alert in the form of a green bar (Fig. \ref{fig:tracking}(b)) and an auditory recorded warning of ``Clear'' were delivered at the start of a new trial.


\textbf{\textit{Toggle between two displays}}. 
Every trial started on the tracking display. Participants were allowed access to only one display at a time and were required to switch between the tracking and detection displays using a ``Switch'' button on the joystick. A 0.5-second time delay was introduced each time they toggled between displays, simulating computer processing and loading. 

\textbf{\textit{Scoring system}}. 
During the task, participants performed both tracking and detection tasks simultaneously, making trade-offs on which to prioritize. 
Each trial offered up to 10 points for tracking and 5 points for detection, with a total of 15 points possible. The ratio between the tracking and detection tasks was determined based on a prior study \cite{du_not_2020}, which found that this ratio ensured the potential gain from focusing on one task was equivalent to the opportunity cost of losing points on the other task. This setup was designed to encourage participants to make trade-off decisions at any given time.

The tracking score was based on a 10-bin histogram of the Root Mean Square of the tracking Errors (RMSE) \cite{yang2017evaluating}, calculated as $\sqrt{\frac{1}{n}\Sigma_{i=1}^{n}{(Tracking \ Error)^2}}$, where $n=200$. Tracking error, measured in pixels at 20 Hz, was converted into a score (0-10), with smaller errors yielding higher scores. The detection score was calculated based on accuracy and response time (see Equation \ref{eq:score}). After each trial and right before answering the trust score for that trial, participants received feedback that presented information including true state, whether the automated threat detector and their identifications were correct, and scores.


\vspace{-10pt}
\begin{small}
\begin{equation}
\label{eq:score}
score =  \\
    \begin{cases}0 & Wrong \\5 - 5 \times\ \frac{Detection \ Time}{10000 \ milliseconds} & Correct \ (threat) \\5 & Correct\ (no \ threat) \end{cases}
\end{equation}
\end{small}

\subsection{Experimental design}
In this study, we manipulated the reliability of the automated threat detector to examine whether the distribution of trust dynamics types varies across different reliability levels. Prior literature suggests that 70\% reliability is considered a threshold at which automation becomes useful \cite{wickens2007benefits}, while reliability below this threshold results in detrimental effects on overall performance \cite{rovira2010transitioning}. Accordingly, prior studies exploring trust dynamics types have focused on reliability levels above this threshold, ranging from 70\% to 90\% \cite{Bhat2022_RAL, guo2021modeling}. In contrast, this study aimed to explore reliability levels slightly below the threshold, ranging from 62\% to 70\%, to investigate how trust dynamics would differ. Participants were randomly assigned to one of the five levels.

The numbers of hits, misses, false alarms (FAs), and correct rejections (CRs) for each reliability level were configured using the Signal Detection Theory \cite{stanislaw1999calculation}. The criterion $c$ and the sensitivity $d'$ values were set at $-0.20$ and $1.09$, respectively, by benchmarking prior literature \cite{manzey2014decision}. Using the predefined values of $c$, $d'$, and the corresponding reliability level, the number of hits, misses, FAs, and CRs were calculated and rounded to the nearest integers (Table \ref{table: SDT number}). 

Once a participant was assigned to a reliability level, the number of hits, misses, FAs, and CRs was determined following Table \ref{table: SDT number}. The sequence of occurrence of hits, misses, FAs, and CRs was randomized. For example, if a participant is assigned to the 62\% reliability condition, s/he will experience 8 hits, 2 misses, 36 FAs, and 54 CRs, following a randomly generated sequence of occurrence.

Each set of four images was randomly selected from a large pool of images that included both those with and without a threat. For example,  when a ``miss'' trial occurs, the set of four images will consist of three images without a threat and one image with a threat (Fig. \ref{fig:detection}). However, the automated threat detector will indicate clear. Threats were randomly distributed across the four images, following a uniform distribution.

\begin{small}
\begin{table}[!ht]
    \caption{Occurrences of hits, misses, false alarms, and correct rejections in each reliability level}
    \label{table: SDT number}
    \centering
    \begin{tabular}{ccccc}
        \toprule
        Reliability & \makecell{Hit \\(Alert: Y \\ Threat: Y)} & \makecell{Miss \\ (Alert: N \\ Threat: Y)} & \makecell{FA \\ (Alert: Y \\ Threat: N)} & \makecell{CR \\ (Alert: N \\ Threat: N)} \\
        \midrule
        \texttt{62\%} & 8 & 2 & 36 & 54 \\
        \texttt{64\%} & 16 & 4 & 32 & 48 \\
        \texttt{66\%} & 24 & 6 & 28 & 42 \\
        \texttt{68\%} & 32 & 8 & 24 & 36 \\
        \texttt{70\%} & 40 & 10 & 20 & 30\\
        \bottomrule
    \end{tabular}
\end{table}
\end{small}

\subsection{Measures}
A range of subjective, behavioral, and performance data were collected in this study. 

\textbf{\textit{Personal characteristics}}. We administered a pre-experimental paper survey to gather data on participants' personal characteristics across 12 constructs (e.g., cultural dimensions, personality traits) encompassing 28 dimensions (e.g., power distance, masculinity, extraversion, agreeableness). This list was developed by synthesizing key review papers that summarized factors influencing users' \textit{snapshot} trust \cite{hoff2015trust, hancock2021evolving, kaplan2023trust, hancock2011meta, schaefer2016meta}. For each construct, we employed well-established scales (Table \ref{table:anova-pc}). For all relevant survey items, the scale was anchored at the endpoints.

\textbf{\textit{Dynamic trust}}. 
The experiment consisted of 100 trials. After each $i^\text{th}$ trial, participants recorded their subjective trust ratings, $trust_{i}$, using a visual analog scale \cite{manzey2012human, du_not_2020, yang_trust_2023} that appeared on the computer screen. The scale had ``I don't trust the detector at all'' as the leftmost anchor to ``I absolutely trust the detector'' as the rightmost anchor. Participants' ratings were then converted and recorded to values between 0 and 1. The self-reported trust ratings, $trust_{i}$, are used as the ground truth labels for modeling and clustering trust dynamics as illustrated in Section \ref{sec:trustdynamics}.

\textbf{\textit{Blindly-following and cross-checking behaviors}}.
In each trial, participants could blindly follow the recommendation provided by the automated threat detector—either by reporting a threat after receiving the detector's ``Danger'' alert or by taking no action following the ``Clear'' signal—without cross-checking the detection display. Alternatively, they could cross-check the detection display themselves to verify the automated threat detector's recommendation. These behaviors were categorized as blindly-following and cross-checking, respectively. We calculated each participant's overall ratios of blindly-following and cross-checking across the 100 trials.




\textbf{\textit{Performance}}. 
Participants' performance scores for the tracking and threat detection tasks were captured. Tracking and detection scores were calculated using the method outlined in Section III-B. The total score was obtained by summing these two scores.

\textbf{\textit{Post-experimental survey}}.
After completing the experiment, through a paper survey, we measured participants' overall trust \cite{uggirala2004measurement}, satisfaction \cite{van1997simple}, and understanding \cite{madsen2000measuring} of the detector, as well as self-confidence \cite{de2003effects} regarding their performance. For all items, the scale was anchored at the endpoints of the Likert scale.


\subsection{Experimental procedure}
Upon arrival, participants provided informed consent and completed a pre-experimental survey on demographics and personal characteristics. They then received instructions and a practice session for the task. The instructions included examples of identifying threats in images with and without threats. Participants were given time to scan four images and decide if a threat was present. The practice session involved 30 trials of the tracking task alone, followed by 8 trials of both tracking and detection tasks. These 8 trials covered all possible outcomes (hits, misses, false alarms, and correct rejections). 
Participants were informed that the automated threat detector used during the practice was for illustration purposes. It was emphasized before and after the practice session that the detector's reliability would be different in the actual experimental trials.

The experiment included 100 trials. After completing a block of 50 trials, which lasted about 12–13 minutes, participants took a five-minute break before proceeding to the second block of 50 trials. Consequently, it took approximately 30 minutes to complete all 100 trials. After completing all the trials, participants filled out a post-experimental survey and were compensated for their participation.

\section{DATA ANALYSIS AND RESULTS}

We conducted k-means clustering on participants' trust dynamics data and identified three distinct types of trust dynamics. Following this, we performed multiple one-way ANOVAs to assess differences among the cluster groups. Finally, we developed a predictive model.

\subsection{Modeling and clustering trust dynamics}\label{sec:trustdynamics}

We model trust as a Beta random variable following a Beta distribution. The shape of the distribution is specified by a Beta probability density function (PDF) \cite{guo2021modeling, yang_trust_2023}.
The Beta distribution has been used in the literature to model reputation management in e-commerce \cite{josang2002beta, josang2007dirichlet}. Models employing the Beta distribution have been shown to outperform other models \cite{lee1992trust, xu2015optimo}. Additionally, such models offer good model explainability and generalizability, as they align with three key properties of trust dynamics, namely continuity, negativity bias, and stabilization, identified in empirical studies \cite{yang_trust_2023}. 




We assume that an individual’s level of trust, after the automated threat detector completes the $i^\text{th}$ task, has a probability distribution that may be represented by the Beta PDF:

\begin{small}
\begin{equation}
    \label{eq:beta}
    t_{i} \sim Beta(\alpha_{i}, \beta_{i})
\end{equation}
\end{small}

The parameters of the Beta distribution, $\alpha_{i}$ and $\beta_{i}$, define the shape of the individual's trust-probability distribution. We assume that the expected mean value of Equation \ref{eq:beta} is the most likely prediction of the individual's trust,  which is given by Equation \ref{eq:betaavg}:

\begin{small}
\begin{equation}
    \label{eq:betaavg}
    \hat{t_{i}} = E(t_{i}) = \frac{\alpha_{i}}{\alpha_{i}+\beta_{i}}
\end{equation}
\end{small}

The Beta distribution parameters $\alpha_{i}$ and $\beta_{i}$ are calculated using Equations \ref{eq:alphaupdate} and \ref{eq:betaupdate}, where $\omega^{s}$ and $\omega^{f}$ represent gains due to the human agent's positive and negative experiences with the threat detector.


\begin{small}
\begin{equation}
    \label{eq:alphaupdate}
    \alpha_{i} = \begin{cases}
        \alpha_{i-1} + \omega^{s}, & if \ p_{i} = 1 \ (threat \ detector's \ success)\\
        \alpha_{i-1}, & if \ p_{i} = 0 \ (threat \ detector's \ failure)
    \end{cases}
\end{equation}

\begin{equation}
    \label{eq:betaupdate}
    \beta_{i} = \begin{cases}
        \beta_{i-1} + \omega^{f}, & if \ p_{i} = 0 \ (threat \ detector's \ failure)\\
        \beta_{i-1}, & if \ p_{i} = 1 \ (threat \ detector's \ success)
    \end{cases}
\end{equation}

where
\begin{itemize}
    \item $p_{i}$: Performance of the automated threat detector on the $i^\text{th}$ task
    \item $\alpha_{i}$, $\beta_{i}$: Parameters of the Beta distribution after the $i^\text{th}$ task
    \item $\omega^{s}$: Gains due to the human's positive experience with the automated threat detector after its success
    \item $\omega^{f}$: Gains due to the human's negative experience with the automated threat detector after its failure
\end{itemize}
\end{small}

After $i$ tasks, the automated threat detector succeeds in $n^s$ tasks and fails $n^f$ tasks ($i=n^s+n^f$). Then:
\vspace{1mm}
\begin{small}
\begin{equation}
    \label{eq:stabilization}
    t_i \sim Beta(\alpha_0+n^sw^s,\beta_0+n^fw^f)
\end{equation}



\begin{equation}
    \label{eq:stab_predict}
    \hat{t_{i}} = E(t_{i}) = \frac{\alpha_0+n^sw^s}{\alpha_0+n^sw^s+\beta_0+n^fw^f}
\end{equation}

where 
\begin{itemize}
    \item $\alpha_0$, $\beta_0$: Human's \textit{a priori} positive and negative experience with automation in general
    \item $\omega^{s}, \omega^{s}$: Gains due to the human's positive and negative experience with the automated threat detector
    \item $n^s, n^f$: number of successes and failures contained in the $i$ tasks
\end{itemize}
\end{small}

In this study, we personalized the trust model for each participant. 
After every trial, we iteratively updated the parameter set $\theta = \left\{\alpha _{0} ,\beta _{0} ,w^{s} ,w^{f}\right\}$ over the 100 trials using the maximum a posteriori estimation (MAP). To determine the initial priors for MAP, we followed the approach used in \cite{guo2021modeling}: We treated one of the 130 participants as a \textit{\textbf{new}} human agent, whose trust needs to be estimated. The remaining 129 participants were regarded as \textit{\textbf{old}} human agents. We assume that before the \textit{\textbf{new}} human agent performed the task, $k$ other \textit{\textbf{old}} human agents had worked with the automated threat detector, and each of the old human agents finished $n$ tasks. Each old human agent reported his or her trust at the end of each task, so his or her trust history $T^j=\{t^j_1,...,t^j_n\}$ and the automated threat detector's performance history $P^j=\{p^j_1,...,p^j_n\}$ are fully available, $j=1,2,...,k$.

This prior can be estimated by the empirical distribution of the parameters of the $k$ old human agents who have previously worked with the same automated threat detector. The parameter $\theta_j$ of human agent $j$ is estimated via the Maximum Likelihood Estimation (MLE):
\begin{equation}
\begin{aligned}
\theta_j  & =\underset{\theta }{\text{argmax}} \ P( T^j\ |\ \theta,P^j )\\
 & =\underset{\theta }{\text{argmax}} \ \prod\limits ^{n}_{i=1} Beta( t^j_{i} ;\alpha^j _{i} ,\beta^j _{i})
\end{aligned}
\label{eq:trainingMLE}
\end{equation}
where $\alpha^j_i$, $\beta^j_i$, $i=1,2,...$, are determined by Equations ~\eqref{eq:alphaupdate} and \eqref{eq:betaupdate}.

The objective is to predict the \textit{\textbf{new}} human agent's trust $t_m$ after s/he finishes the $m^\text{th}$ task, based on the automated threat detector's performance history $P_m=\{p_i|i=1,2,3,...,m\}$, reported trust $T_m^o=\{t_i|i=1,2,3,...,m-1\}$, and the data $T^j$ and $P^j$ from the $k$ old human agents, $j=1,2,...,k$.

Personalizing the trust model for the new human agent means finding the best $\theta$ for him or her. Here, we use the maximum a posteriori estimation (MAP) to estimate $\theta$, which is to maximize the posterior of $\theta$, given the automated threat detector's performance $P_m$ and trust history $T_m^o$. First, we have
\begin{align}
\begin{split}
& P( \theta \ |\ P_m,T_m^o)\\
  \propto &\prod _{t_{i} \in T_m^o} Beta( t_{i} ;\alpha _{i} ,\beta _{i}) \ \cdot P( \theta )
\end{split}
\end{align}
Then
\vspace{1mm}
\begin{align}
\begin{split}
    \theta = & \underset{\theta }{\text{argmax}} \ P( \theta \ |\ P_m,T_m)\\
=&\underset{\theta }{\text{argmax}} \prod _{t_{i} \in T_m^o} Beta( t_{i} ;\alpha _{i} ,\beta _{i}) \ \cdot P( \theta )\\
=&\underset{\theta }{\text{argmax}}\sum _{t_{i} \in T_m^o}\log( Beta( t_{i} ;\alpha _{i} ,\beta _{i})) \ +\log P( \theta )
\end{split}
\label{eq:theta_MAP}
\end{align}
where $P( \theta )$ is the prior determined by Equation \eqref{eq:trainingMLE}.

\textbf{\textit{Clustering trust dynamics}}. We applied k-means clustering to classify participants within each reliability level. Following prior studies \cite{guo2021modeling, yang_trust_2023}, the clustering was based on two factors: average logarithm trust and RMSE.

\begin{small}
    
\begin{equation}
    Average \ logarithm \ trust = \frac{1}{100} \sum_{i=1}^{100} \log({t}_{i})
\end{equation}

\begin{equation}
    RMSE = \sqrt{\frac{1}{100} \sum_{i=1}^{100} \left(t_{i} - \hat{t}_{i}\right)^{2}}
\end{equation}

where 
\begin{itemize}
    \item $t_{i}$: Participant's self-reported trust in the automated threat detector after the $i^\text{th}$ task
    \item $\hat{t_{i}}$: Predicted trust value for the $i^\text{th}$ task (Equation \ref{eq:stab_predict})
\end{itemize}
\end{small}


The average logarithm trust was selected to assess how much participants trusted the automated threat detector on average across the 100 trials. The RMSE measures the deviation between the self-reported ground truth and the predicted trust value. The RMSE was chosen as it captures the pattern of trust change. Specifically, this study is interested in evaluating individual trust dynamics to determine whether the pattern of change follows the key properties of trust dynamics mentioned earlier (continuity, negativity bias, stabilization).

\begin{figure}[ht]
  \centering\subfloat[Example scree plot (Optimal number of clusters: 3)]{\includegraphics[scale=0.4]{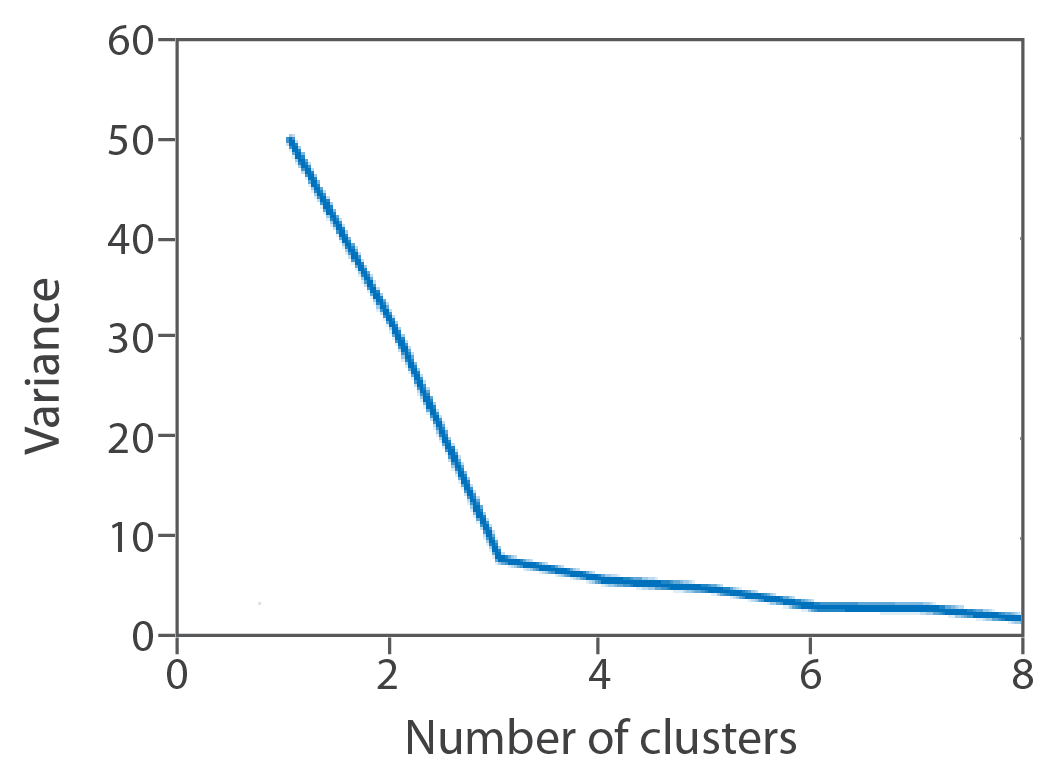}}
  \hspace{0.05 em}
  \subfloat[Distribution of clusters. Three clusters were identified: BDM, disbeliever, and oscillator.]{\includegraphics[scale=0.17]{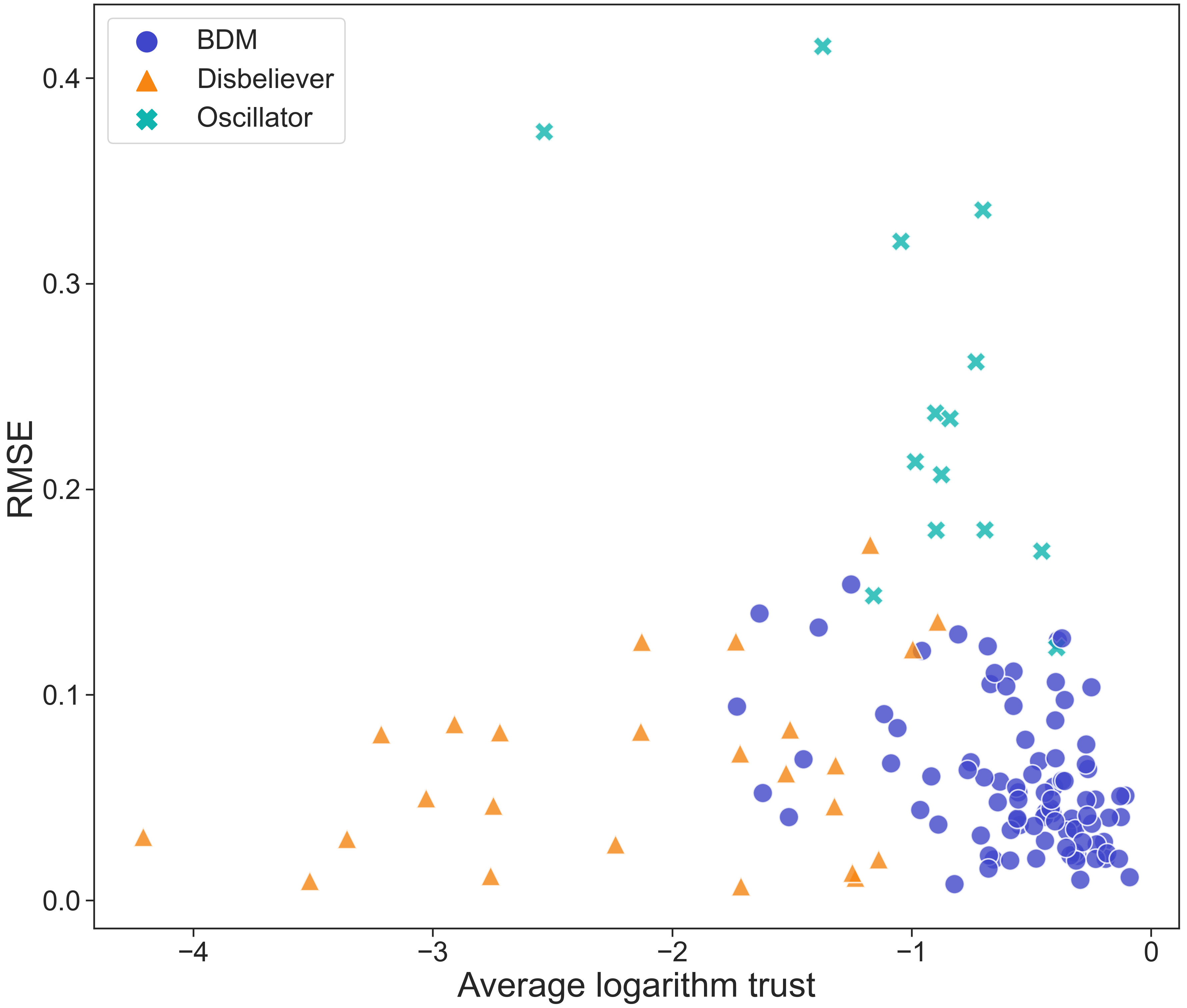}}
  \caption{K-means clustering results. The elbow rule determined the optimal number of clusters to be three (a). Subsequently, three clusters (BDM, disbeliever, and oscillator) were identified (b).}
  \label{fig:clustering}
\end{figure}

\begin{figure*}[!ht]
  \centering
  \includegraphics[scale=0.35]{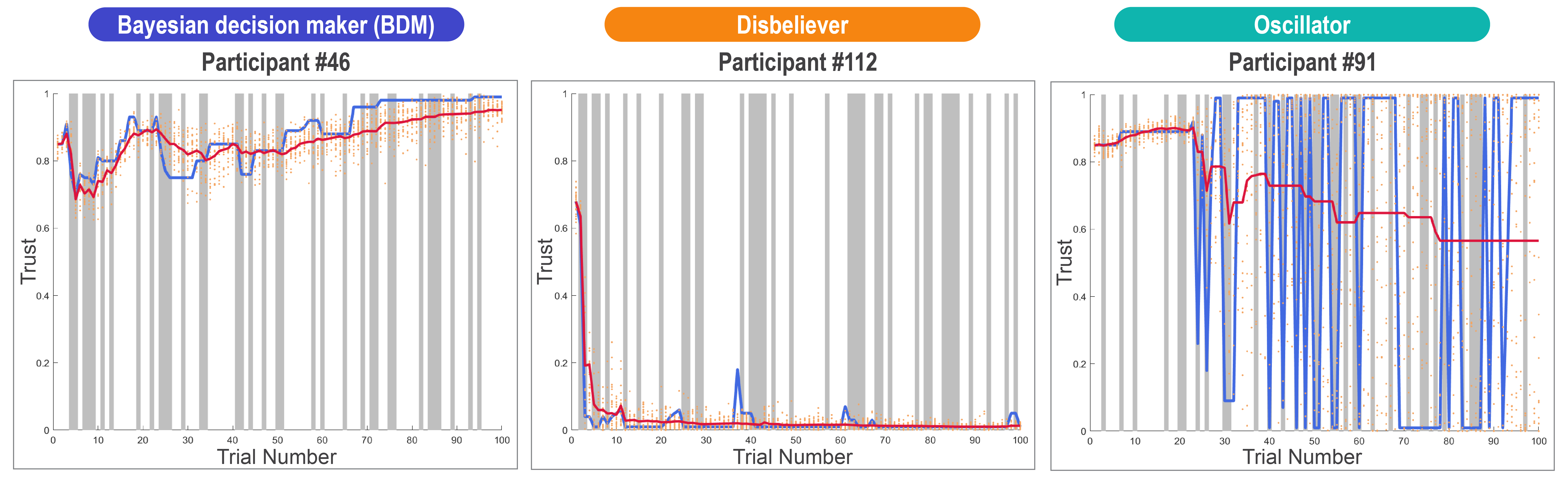}
  \caption{Examples of trust dynamics from each cluster. The BDMs update their trust in a Bayesian manner, the disbelievers display constantly low trust, and the oscillators' trust fluctuates dramatically. (Blue line: Participants' self-reported trust; Red line: Predicted trust $\hat{t_{i}}$; Orange dots: Random samples from $t_{i} \sim Beta(\alpha_{i}, \beta_{i})$ ; Grey columns: Trials where automation failed)}
  \label{fig:3clusters}
\end{figure*}

The elbow rule was used to decide on the optimal number of clusters that best capture the patterns in the data (Fig. \ref{fig:clustering}(a)). Consistent with prior literature \cite{guo2021modeling, Bhat2022_RAL, yang_trust_2023}, three distinct clusters were identified: Bayesian decision makers (BDMs, $n=91$), disbelievers ($n=25$), and oscillators ($n=14$). 
The BDM group showed relatively high average logarithm trust and low RMSE, the disbeliever had low average trust and RMSE, and the oscillator group is characterized by high RMSE (Table \ref{table:clusterstat}, Fig. \ref{fig:clustering}(b)). Fig. \ref{fig:3clusters} illustrates examples of trust dynamics, one from each of the three clusters.
Irrespective of the reliability level, there consistently existed all three types of trust dynamics. We conducted a chi-squared test to determine whether the distribution of clusters differed by reliability level, and no significant difference was found ($\chi^{2}=5.77, df=8, N=130, p=0.673$). As a result, we combined the data from different reliability groups for subsequent analyses.


\begin{small}
\begin{table}[!ht]
    \caption{Average values of the clustering variables and the final parameter set $\theta$}
    \label{table:clusterstat}
    \centering
    \begin{tabular}{ccccccc}
        \toprule
        Cluster & \makecell{Average \\ logarithm \\ trust} & RMSE & $\alpha _{0}$ & $\beta _{0}$ & $w^{s}$ & $w^{f}$ \\
        \midrule
        BDM & -0.554 & 0.057 & 231.93& 128.35& 2.21& 1.88\\
        Disbeliever & -2.099 & 0.064 & 71.78& 335.22 &0.43 &6.48\\
        Oscillator & -0.970 & 0.243 & 3.83 &3.47&0.04&0.07\\
        \hline
        All & -0.896 & 0.078 &176.57&154.68&1.64&2.57 \\
        \bottomrule
    \end{tabular}
\end{table}
\end{small}

\subsection{Associations between trust dynamics, personal characteristics, behaviors, and performance} \label{sec:association}
After the clustering analysis, each participant was assigned to one of the three clusters. We gathered cluster assignment data from the five reliability groups. Subsequently, we conducted one-way ANOVAs to examine differences between the three clusters. The independent variable was the cluster group while the dependent variables included all relevant measures, including personal characteristics, behavior and performance metrics, and post-experimental surveys. In cases where there were significant differences, post-hoc tests with Bonferroni corrections were conducted.

\textbf{\textit{Personal characteristics}}.
Table \ref{table:anova-pc} tabulates the mean and standard deviation (SD) values for the personal characteristics dimensions across the three clusters. 

\begin{small}
\begin{table*}[!ht]
    \caption{Cluster differences in personal characteristics (mean and SD)}
    \label{table:anova-pc}
    \centering  
    \begin{tabular}{clccc}
    \hline
    Construct (Scale) & Dimension                      & BDM          & Disbeliever  & Oscillator  \\
    \hline
    Culture (CVScale \cite{yoo2011measuring}) & Power distance (/5)            & 1.82 (0.46) & 1.80 (0.54)  & 1.86 (0.73) \\
    & Uncertainty avoidance (/5)     & 4.25 (0.48) & 4.36 (0.50)  & 4.27 (0.58) \\
    & Collectivism (/5)              & 3.08 (0.63) & 3.21 (0.69)  & 3.21 (0.90) \\
    & Long-term orientation (/5)     & 4.09 (0.47) & 4.19 (0.52)  & 4.26 (0.36) \\
    & Masculinity (/5) \mbox{*}             & 1.81 (0.67) & 1.68 (0.77)  & 2.38 (1.12) \\
    \hline
    Attentional control (ACS \cite{derryberry2002anxiety}) & Attentional capacity (/4)      & 2.67 (0.40) & 2.58 (0.38)  & 2.79 (0.32) \\
    \hline
    Mood (PANAS \cite{watson1988development}) & Positive affect (/5) \mbox{*}         & 2.80 (0.71) & 2.75 (0.76)  & 3.37 (0.65) \\
    & Negative affect (/5)           & 1.36 (0.45) & 1.35 (0.31)  & 1.31 (0.41) \\
    \hline
    Personality (Mini-IPIP \cite{donnellan2006mini})& Extraversion (/5) \mbox{*}            & 2.99 (0.95) & 2.74 (0.96)  & 3.54 (1.08) \\
    & Agreeableness (/5)             & 3.99 (0.64) & 3.90 (0.75)  & 3.96 (0.70) \\
    & Conscientiousness (/5)         & 3.38 (0.86) & 3.15 (1.17)  & 3.57 (0.58) \\
    & Neuroticism (/5) \mbox{*}             & 2.67 (0.73) & 3.09 (0.83)  & 2.61 (0.90) \\
    & Intellect (/5) \mbox{*}               & 3.77 (0.72) & 3.57 (0.93)  & 4.23 (0.46) \\
    \hline
    Risk propensity (RPS \cite{meertens2008measuring}) & Risk propensity (/9)           & 4.20 (1.28) & 4.15 (1.38)  & 4.83 (1.14) \\
    \hline
    Decision-making style (GDMS \cite{spicer2005examination}) & Intuitive (/5)           & 3.35 (0.65) & 3.23 (0.74)  & 3.53 (0.86) \\
    & Dependent (/5)                 & 3.75 (0.67) & 3.74 (0.68)  & 3.54 (0.65) \\
    & Rational (/5)                  & 4.09 (0.49) & 4.03 (0.81)  & 4.36 (0.42) \\
    & Avoidant (/5)                  & 2.97 (1.04) & 2.88 (1.09)  & 2.43 (1.00) \\
    & Spontaneous (/5)               & 2.65 (0.70) & 2.82 (0.68)  & 2.63 (0.63) \\
    \hline
    Reasoning (CRT \cite{toplak2014assessing}) & Reasoning test score (/7)      & 5.56 (1.42) & 5.04 (1.79)  & 5.14 (2.32) \\
    \hline
    Propensity to trust (PTS \cite{merritt2013trust}) & Trust propensity (/5) \dag        & 3.29 (0.44) & 3.05 (0.69)  & 3.13 (0.64) \\
    \hline
    Negative attitude towards & Attitude-Interaction (/5)      & 2.17 (0.63) & 2.26 (0.68)  & 1.96 (0.53) \\
    autonomous systems (NARS \cite{nomura2006measurement}) & Attitude-Social influence (/5) & 3.22 (0.60) & 3.41 (0.82)  & 3.09 (0.94) \\
    \hline
    Expectancy towards & Performance expectancy (/7) \mbox{**} & 5.73 (0.70) & 5.24 (1.39)  & 6.07 (0.68) \\
    autonomous systems (UTAUT  \cite{venkatesh2003user})  & Effort expectancy (/7)         & 5.24 (0.72) & 5.17 (0.78)  & 5.57 (0.99) \\
    \hline
    Self-efficacy towards autonomous systems (Efficacy beliefs \cite{hill1987role}) & Self-efficacy (/5)             & 3.45 (0.74) & 3.63 (0.88)  & 3.82 (0.81) \\
    \hline
    Perfect automation schema (PAS \cite{merritt2015measuring}) & PAS-High expectations (/5) \mbox{**}   & 1.88 (0.57) & 1.55 (0.54)  & 2.14 (0.79) \\
    & PAS-All or none (/5)           & 1.80 (0.62) & 2.00 (0.87)  & 1.79 (0.70) \\
    \hline
    \end{tabular}%
    \\
    \mbox{***}-$p<0.001$, \mbox{**}-$p<0.01$, \mbox{*}-$p<0.05$, \dag-$p<0.1$
\end{table*}
\end{small}

We found significant differences in seven personal characteristics dimensions: \textit{masculinity} ($F(2, 127)=4.21, p=0.02$), \textit{positive affect} ($F(2, 127)=4.11, p=0.019$), \textit{extraversion} ($F(2, 127)=3.06, p=0.05$), \textit{neuroticism} ($F(2, 127)=3.16, p=0.046$), \textit{intellect} ($F(2, 127)=3.63, p=0.029$), \textit{performance expectancy} ($F(2, 127)=4.76, p=0.01$), and \textit{PAS-high expectations} ($F(2, 127)=5.08, p=0.008$). 

\begin{figure*}[!ht]
    \centering\subfloat[Masculinity]{\includegraphics[scale = 0.6]{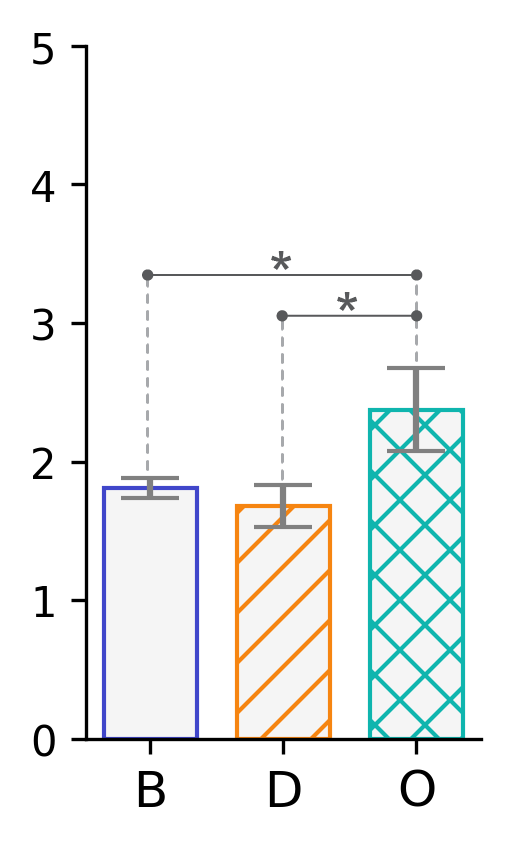}}
    \subfloat[Positive affect]{\includegraphics[scale = 0.6]{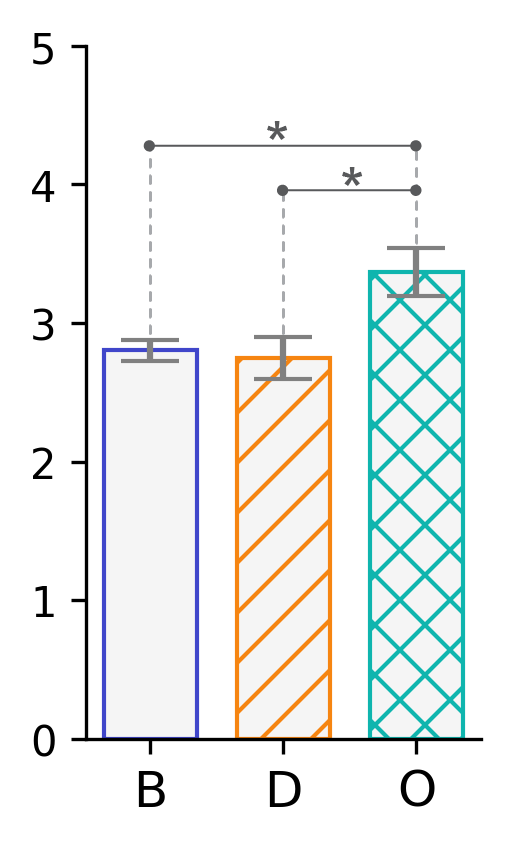}}
    \subfloat[Extraversion]{\includegraphics[scale = 0.6]{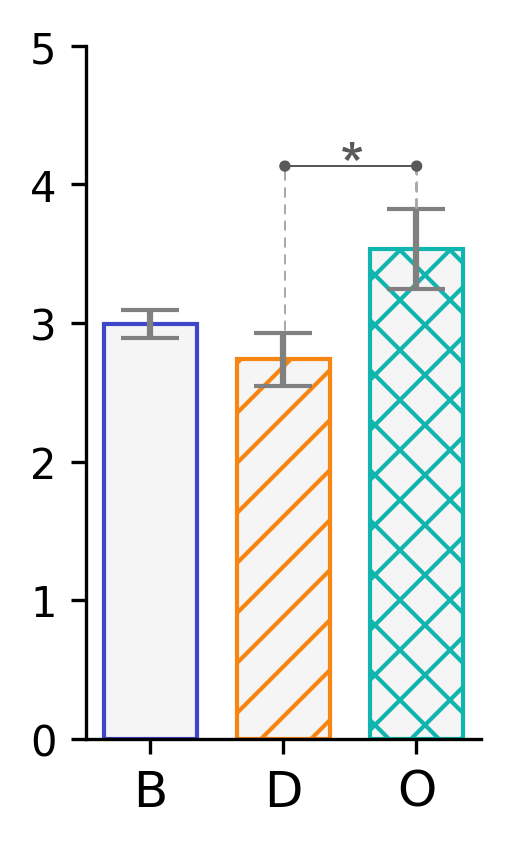}}
    \subfloat[Neuroticism]{\includegraphics[scale = 0.6]{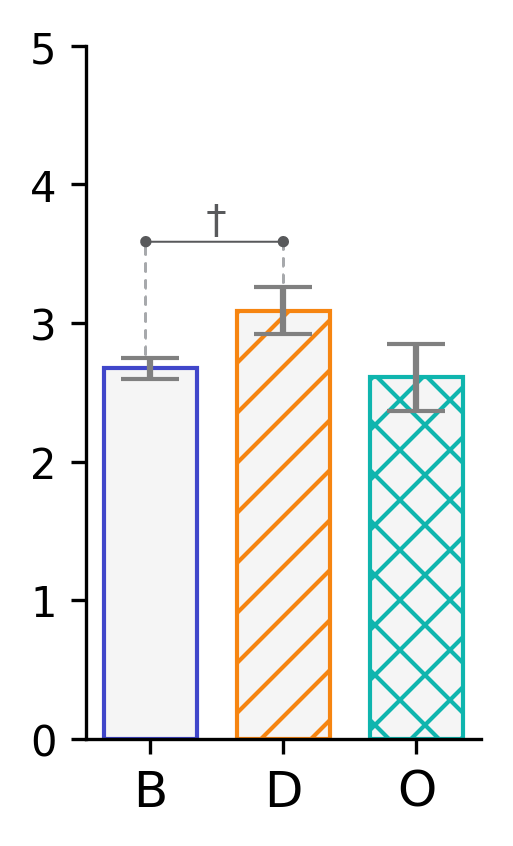}}
    \centering\subfloat[Intellect]{\includegraphics[scale = 0.6]{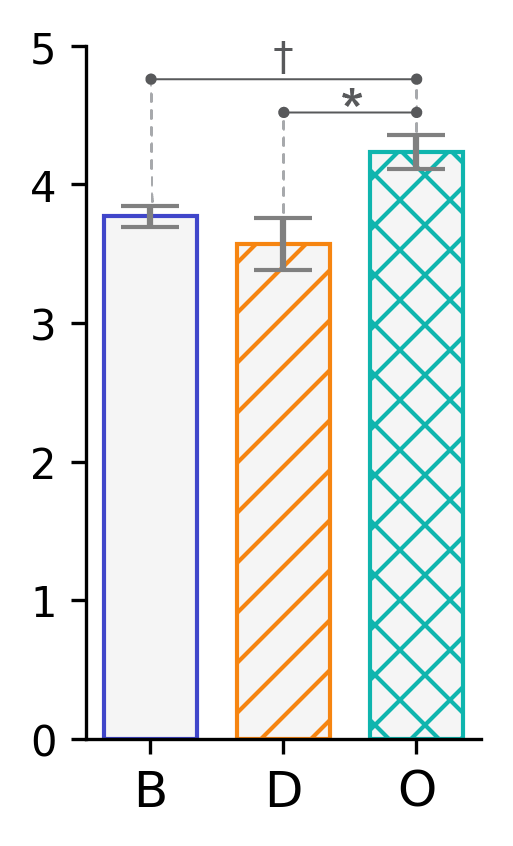}}
    \subfloat[Performance expectancy]{\includegraphics[scale = 0.6]{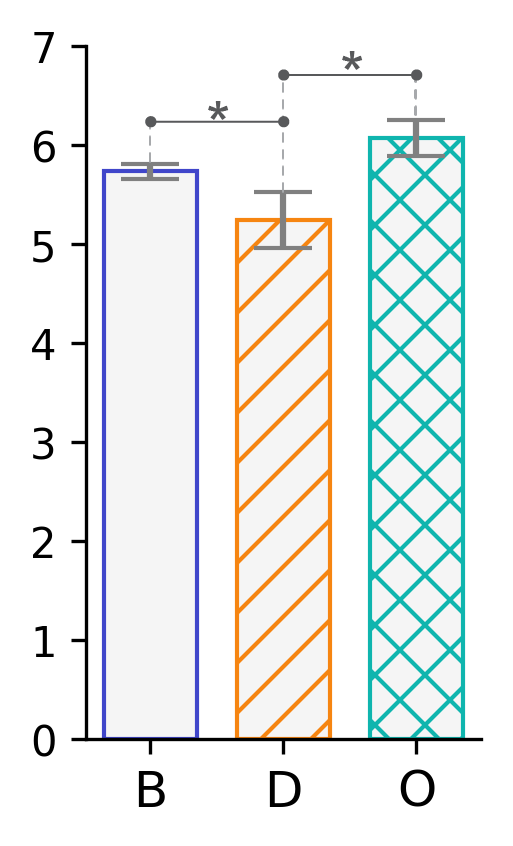}}
    \subfloat[PAS-High expectations]{\includegraphics[scale = 0.6]{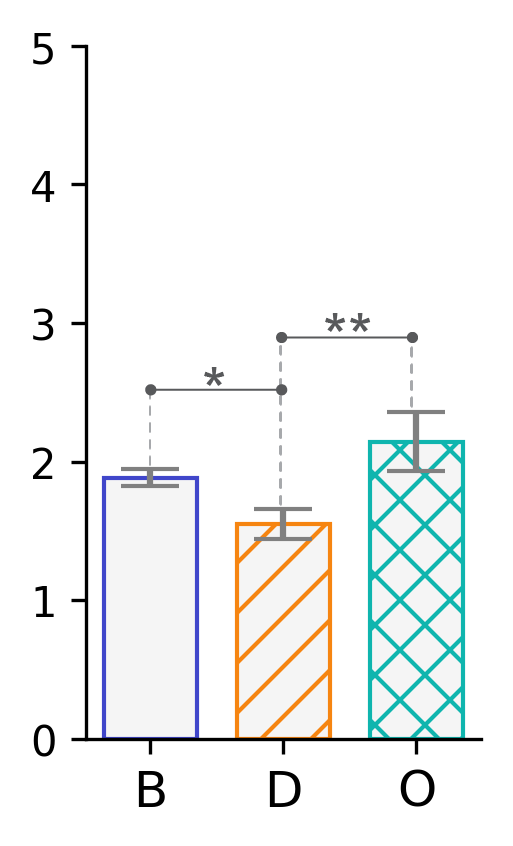}}
        \caption{Mean comparison of personal characteristics between the clusters (BDM (B), disbeliever (D), and oscillator (O)). \mbox{**}-$p<0.01$, \mbox{*}-$p<0.05$, \dag-$p<0.1$}
    \label{fig:personal}
\end{figure*}

Fig. \ref{fig:personal} (a)-(g) show the pairwise comparisons. 
Regarding \textit{masculinity}, also known as gender role differentiation \cite{yoo2011measuring}, oscillators had significantly higher ratings than BDMs ($p=0.029$) and disbelievers ($p=0.019$). Similarly, concerning \textit{positive affect}, the measure of positive emotions (i.e., cheerfulness, enthusiasm, energy) \cite{watson1988development}, oscillators had significantly higher scores than BDMs ($p=0.021$) and disbelievers ($p=0.032$). With respect to \textit{extraversion}, a personality trait characterized by being outgoing and energetic \cite{donnellan2006mini}, oscillators had higher ratings than disbelievers ($p=0.045$).

\textit{Neuroticism} refers to an individual's tendency to experience negative emotions such as anxiety, anger, fear, sadness, and frustration \cite{donnellan2006mini}. We found that the disbeliever group had the highest average score, and its difference between that of BDM group showed marginal significance ($p=0.053$). Regarding \textit{intellect}, a personality trait of having wide interests and being imaginative \cite{donnellan2006mini}, the average score of oscillators was marginally larger than that of BDMs ($p=0.095$) and disbelievers ($p=0.025$).


Concerning \textit{performance expectancy}, the extent to which a person believes that utilizing the autonomous system would lead to improvements in their work performance \cite{venkatesh2003user}, disbelievers had significantly lower scores compared to BDMs ($p=0.041)$ and oscillators ($p=0.015$). Likewise, regarding \textit{PAS (Perfect Automation Schema)}-High expectations, which measures the extent to which individuals believe that the automated aid will perform with near-perfect reliability \cite{merritt2015measuring}, disbelievers had lower scores compared to BDMs ($p=0.041$) and oscillators ($p=0.010)$.


The three clusters also demonstrated differences in behaviors, performance scores, and post-experimental subjective ratings (Table \ref{table:anova-others}).

\begin{small}
\begin{table*}[!ht]
    \caption{Cluster differences in behaviors, performance, and post-experimental subjective ratings (mean and SD)}
    \label{table:anova-others}
    \centering  
    \begin{tabular}{clccc}
    \hline
    Construct & Dimension                      & BDM          & Disbeliever  & Oscillator  \\
    \hline
    Behavioral metrics & Blindly-following (/1) \mbox{***}     & 0.42 (0.30) & 0.18 (0.22)  & 0.40 (0.22) \\
    & Cross-checking (/1) \mbox{**}          & 0.51 (0.35) & 0.78 (0.28)  & 0.54 (0.27) \\
    \hline
    Performance metrics & Tracking score (/1000) \mbox{**} & 695.14 (188.32) & 561.72 (216.93) & 712.00 (151.15) \\
    & Detection score (/500) & 387.25 (51.57) & 401.51 (57.52) & 374.52 (39.27) \\
    & Total score (/1500) \mbox{*} & 1082.39 (174.60) & 963.23 (179.96) & 1086.52 (150.94) \\
    \hline
    Post-experimental & Trust (/5) \mbox{***}                 & 3.18 (0.60) & 2.26 (0.43)  & 2.88 (0.69) \\
    subjective ratings & Satisfaction (/5) \mbox{***}          & 3.66 (0.73) & 2.67 (0.59)  & 3.31 (0.65) \\
    & Understanding (/5) \mbox{*}           & 3.83 (0.66) & 3.36 (0.89)  & 3.83 (0.48) \\
    & Self-confidence (/7) \mbox{*}         & 4.07 (1.61) & 5.12 (1.54)  & 4.50 (1.70) \\
    \hline
    \end{tabular}%
    \\
    \mbox{***}-$p<0.001$, \mbox{**}-$p<0.01$, \mbox{*}-$p<0.05$, \dag-$p<0.1$
\end{table*}
\end{small}

\begin{figure}[!ht]
    \centering\subfloat[Blindly-following]{\includegraphics[scale = 0.6]{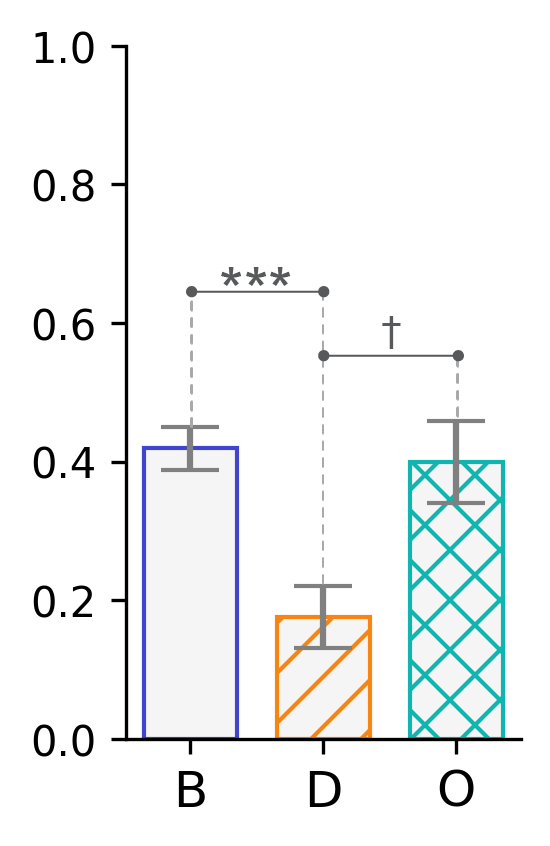}}
    \hspace{0.25 em}
    \subfloat[Cross-checking]{\includegraphics[scale = 0.6]{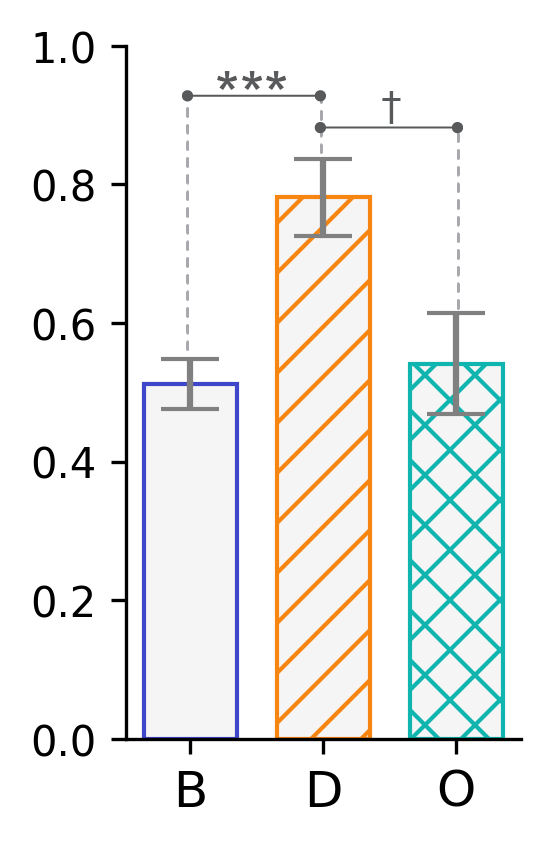}}
    \caption{Mean comparison of the behavior metrics between the clusters (BDM (B), disbeliever (D), and oscillator (O)). \mbox{***}-$p<0.001$, \dag-$p<0.1$}
    \label{fig:behavior}
\end{figure}

\textbf{\textit{Behaviors}}. 
There were significant differences in participants' \textit{blindly-following} ($F(2, 127)= 7.64, p<0.001$) and \textit{cross-checking} ($F(2, 127)= 6.71, p = 0.002$) behaviors. 
The disbeliever group exhibited a lower ratio of blindly-following the automated threat detector compared to the BDM ($p<0.001$) and oscillator groups ($p=0.051$). Conversely, the disbeliever group showed the highest ratio of cross-checking than both the BDM ($p=0.001$) and oscillator ($p=0.089$) (Fig. \ref{fig:behavior}).

\begin{figure}[!ht]
    \centering\subfloat[Tracking score]{\includegraphics[scale = 0.6]{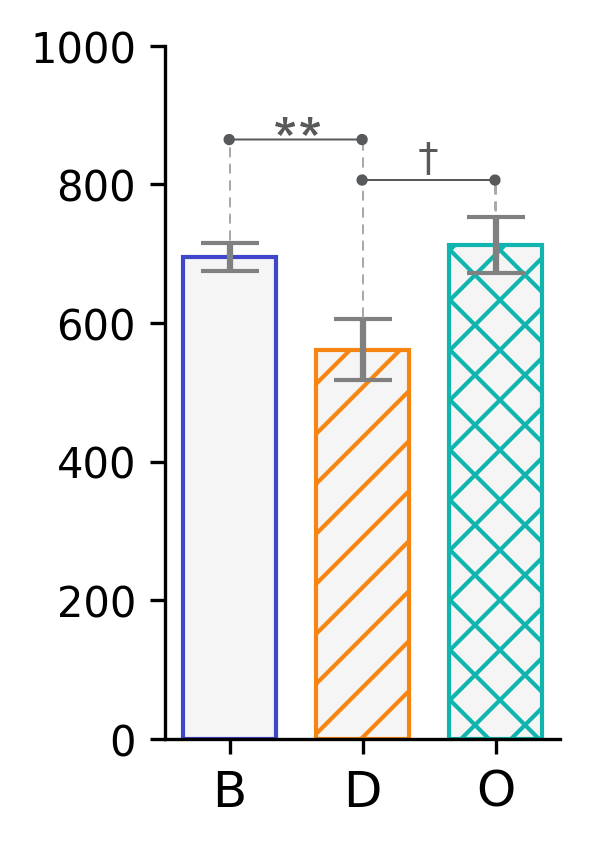}}
    \hspace{0.25 em}
    \subfloat[Detection score]{\includegraphics[scale = 0.6]{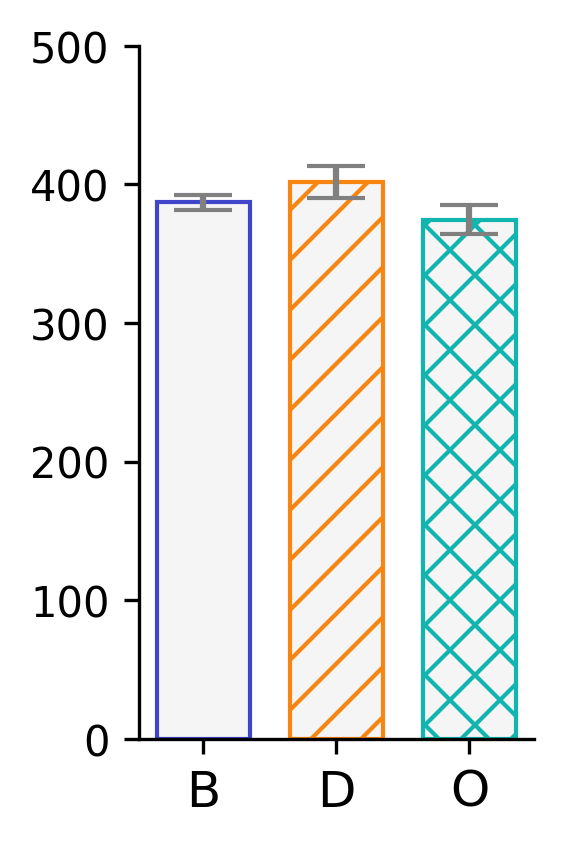}}
    \caption{Mean comparison of the performance metrics between the clusters (BDM (B), disbeliever (D), and oscillator (O)). \mbox{***}-$p<0.001$, \mbox{*}-$p<0.05$}
    \label{fig:performance}
\end{figure}

\textbf{\textit{Performance}}. There were significant differences in \textit{tracking score} ($F(2, 127)= 5.158, p = 0.007$) and \textit{total score} ($F(2, 127)= 4.818, p = 0.010$) among the three clusters. The disbeliever group showed a significantly lower tracking score compared to the BDM ($p=0.007$) and oscillator groups ($p=0.059$). Their total score was also lower than the BDM group ($p=0.009$). On the other hand, the detection scores of the three groups were not significantly different (Fig. \ref{fig:performance}).

\begin{figure}[!ht]
    \centering\subfloat[Trust]{\includegraphics[scale = 0.6]{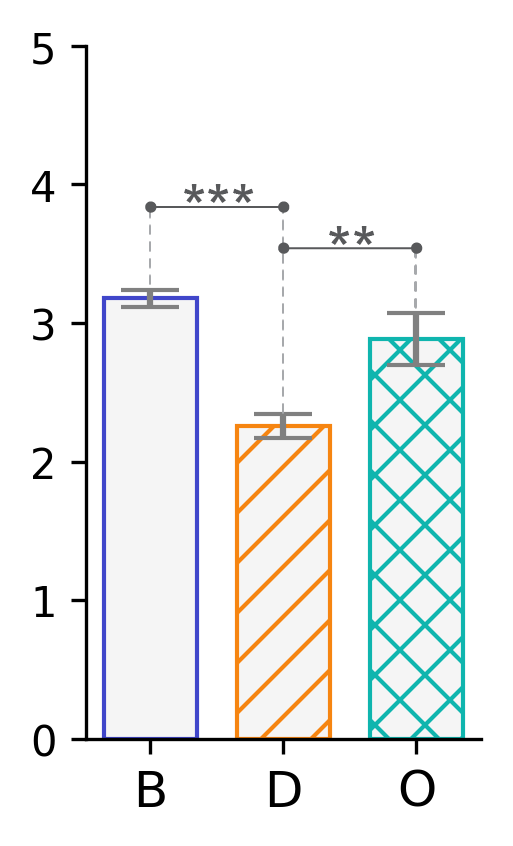}}
    \hspace{0.25 em}
    \subfloat[Satisfaction]{\includegraphics[scale = 0.6]{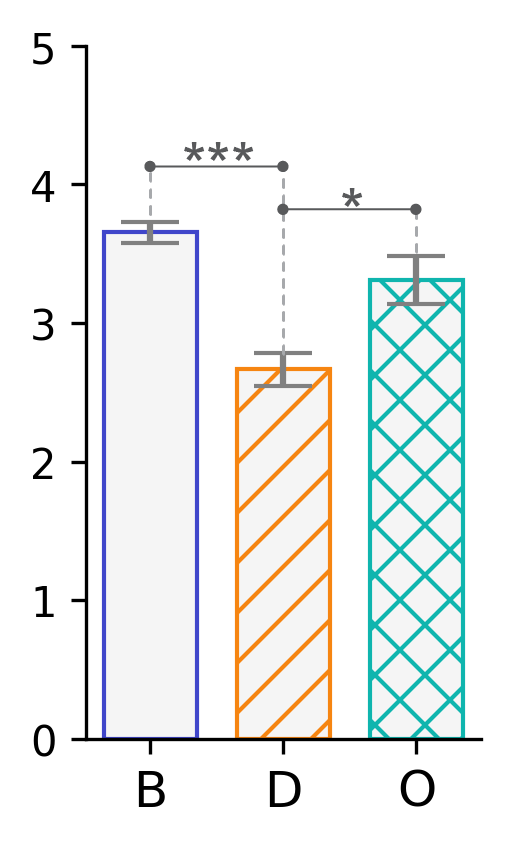}}
    \caption{Mean comparison of the key post-experimental subjective ratings between the clusters (BDM (B), disbeliever (D), and oscillator (O)). \mbox{***}-$p<0.001$, \mbox{**}-$p<0.01$, \mbox{*}-$p<0.05$}
    \label{fig:post}
\end{figure}

\textbf{\textit{Post-experimental subjective ratings}}. The three clusters showed significant differences in the post-experimental subjective ratings: \textit{post-experimental trust} ($F(2, 127)=25.07, p<0.001$), \textit{satisfaction} ($F(2, 127)=20.08, p<0.001$), \textit{understanding} ($F(2, 127)=4.60, p=0.012$), and \textit{self-confidence} ($F(2, 127)=4.33, p=0.015$). In terms of \textit{post-experimental trust}, disbelievers showed significantly lower scores compared to both BDMs ($p<0.001$) and oscillators ($p=0.004$) (Fig. \ref{fig:post}(a)). Similarly, for \textit{satisfaction}, which assessed the participant's satisfaction with the automated threat detector in terms of its usefulness, enjoyability, and effectiveness, the differences were significant between disbelievers and BDMs ($p<0.001$) and oscillators ($p=0.019)$ (Fig. \ref{fig:post}(b)). Concerning \textit{understanding}, which pertains to the degree to which participants could create a mental model and anticipate the behavior of the automated threat detector, disbelievers had significantly lower scores compared to BDMs ($p=0.010$). Finally, concerning \textit{self-confidence}, the disbelievers showed significantly higher scores than BDMs ($p=0.013$).

\subsection{Decision tree model}
To predict the type of trust dynamics a user would exhibit based on their key personal characteristics, we developed a decision tree model using the seven dimensions (\textit{masculinity, positive affect, extraversion, neuroticism, intellect, performance expectancy, PAS-high expectations}) identified in Section \ref{sec:association}.

We randomly split the dataset into training and test sets in an 80:20 ratio. Within the training set, we conducted hyperparameter tuning using a five-fold cross-validation method. For each fold, we trained on 80\% of the training dataset (64\% of the total data points) and validated on the remaining 20\% (16\% of the total data points). Our goal was to identify the parameters (e.g., maximum tree depth, minimum number of leaves) that would yield the highest average weighted F1 score for the validation datasets. We selected the weighted F1 score as the primary criterion, given the imbalanced distribution of the cluster groups.

\begin{figure}[!ht]
  \centering
  \includegraphics[scale=0.44]{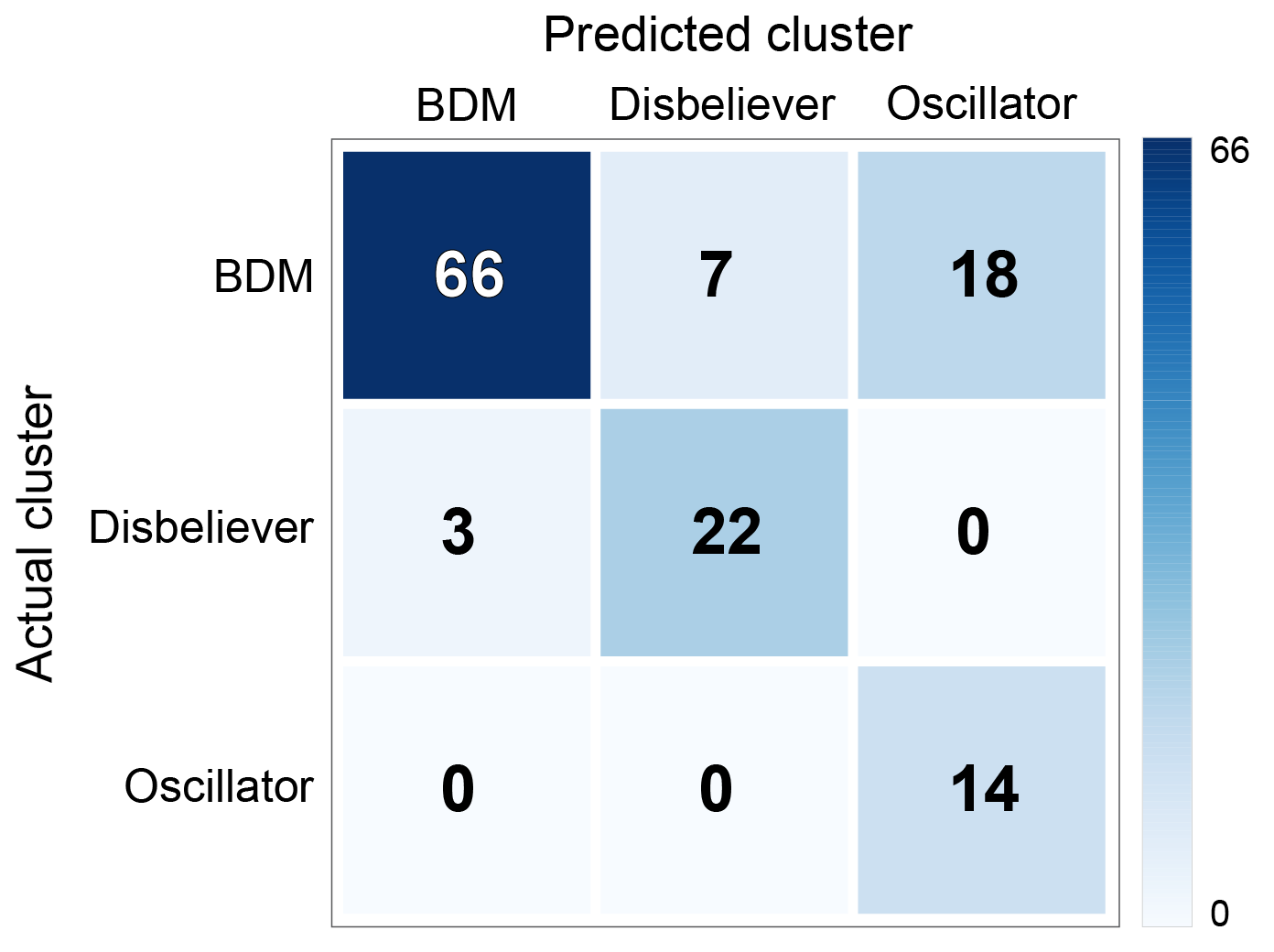}
  \caption{Confusion matrix of the decision tree predicting trust dynamics types based on seven dimensions of personal characteristics}
  \label{fig:confusion}
\end{figure}

After cross-validation, we determined the optimal hyperparameters: a maximum depth of 6, a minimum of 1 leaf per node, and the entropy splitting algorithm. Using these hyperparameters, we configured a decision tree model with the entire training dataset and then tested it with the test dataset.
The tree achieved an accuracy of 70.0\% and demonstrated fairly good performance, particularly in terms of recalling disbelievers and oscillators (Fig. \ref{fig:confusion}).

\section{DISCUSSION}


\subsection{Clustering trust dynamics}
This study focused on exploring trust dynamics using a large participant sample ($n = 130$). We employed the Beta distribution to model participants' real-time trust and identified three distinct trust dynamics types: BDM, disbeliever, and oscillator.  
Notably, the clustering results showed similarities with previous works \cite{Bhat2022_RAL, guo2021modeling}. With a larger sample size ($n=130$), the current study further supports the generalizability of the three trust dynamics types across multiple datasets, even when automation reliability falls below 70\%, ranging from 62\% to 70\%.

\subsection{Associations between trust dynamics and personal characteristics}

We examined an array of personal characteristic variables known to influence (snapshot) trust and identified meaningful associations between individuals' personal characteristics and trust dynamics. Seven personal characteristics dimensions differed significantly among the three clusters: \textit{masculinity}, \textit{positive affect}, \textit{extraversion}, \textit{neuroticism}, \textit{intellect}, \textit{performance expectancy}, and \textit{PAS-high expectations}. The results particularly highlight distinct characteristics of the disbeliever and oscillator groups.

The disbelievers are characterized by high scores for \textit{neuroticism} while low scores for \textit{performance expectancy} and \textit{PAS-high expectations}. In prior research, the \textit{neuroticism} personality trait has been shown to have a negative relationship with post-experimental trust \cite{zhou2020effects, sharan2020effects}. In addition, individuals with high \textit{neuroticism} tend to rely more on their own judgment instead of on automation \cite{zhou2020effects}. The alignment between our results and prior research supports a coherent picture that disbelievers, being highly neurotic, have the lowest post-experimental trust and the highest ratio of cross-checking the automated threat detector. The consistency also provides validity for clustering the disbelievers as a distinct group. The \textit{performance expectancy} and \textit{PAS-high expectations} dimensions both measure an individual's performance expectation of automated/autonomous technologies \cite{venkatesh2003user, merritt2015measuring}. Prior research suggests positive relationships between expectancy and trust \cite{chien2014empirical}, as well as between the high expectations facet of PAS and trust \cite{lyons2019individual}. Similarly, our study suggests that disbelievers have low prior expectations of automation, making it likely that their trust remains low throughout the interaction.

Regarding the oscillator group, they displayed higher scores in \textit{masculinity}, \textit{positive affect}, \textit{extraversion}, and \textit{intellect}. Compared to previous research studies, these results have interesting implications. In fact, \textit{positive affect} \cite{stokes2010accounting}, \textit{extraversion} \cite{Merritt:2008ds, zhou2020effects}, and \textit{intellect} \cite{schaefer2016will} are factors that have been proposed to have a positive correlation with trust. On the other hand, the cultural dimension of \textit{masculinity} has been suggested to make the transfer of trust across entities difficult \cite{doney1998understanding}. However, in our study, all of these dimensions were more prominent for the oscillator group, whose trust neither remained consistently high nor low but fluctuated significantly. 
Still, since the number of oscillators in this study was 14, further research focused on exploring the characteristics of oscillators could greatly enhance our understanding of this group.

\subsection{Associations between trust dynamics, behaviors, performance, and post-experimental subjective ratings}

Our results indicate that the three clusters exhibited significantly different behavioral patterns when interacting with the automated threat detector. These results also help explain the variations in performance scores and post-experimental ratings. Notably, disbelievers showed a lower ratio of blindly-following, lower tracking and total scores, as well as lower post-experimental trust, satisfaction, and understanding than the other two groups.

The disbelievers were less likely to follow the automated threat detector blindly; instead, for most of the trials, they cross-checked and made decisions after spending time scanning the four photos. This distinctive behavior could explain their lower tracking scores compared to the other two groups. Their choice to cross-check might have caused them to lose points on the tracking task, as the crosshair continued to float across the screen regardless of which task (tracking or detection) they focused on.

In contrast, no significant difference was found in the detection score across the clusters. It is worth noting that the detection score was calculated based on both accuracy (whether participants correctly identified the threat or did nothing in the absence of a threat) and speed (how quickly they identified the threat). Disbelievers, who spent more time scanning the four photos themselves, were likely more accurate than those who blindly followed the automated threat detector, which had a reliability of 62-70\%. However, disbelievers likely took more time before pressing the joystick button, as they needed to review the photos. As a result, their reaction time for reporting may have been slower than those who blindly followed the automation. The latter group may have pressed the button immediately after perceiving the alert at the beginning of the trial. Consequently, disbelievers likely lost some points due to slower reaction times. This trade-off between accuracy and speed may have contributed to the lack of a significant difference in detection scores.

Lastly, regarding the post-experimental subjective ratings, it is understandable that the disbelievers, who rarely followed the automation blindly, had the lowest ratings for trust, satisfaction, and understanding. In contrast, they reported higher self-confidence ratings probably because they invested more deliberate effort in the detection task by cross-checking.

\subsection{Predicting trust dynamics using personal characteristics}
We developed a decision tree model to predict the type of trust dynamics an individual might exhibit based on seven personal characteristics. This model can be used to predict the type of trust dynamics a person may display before any real interaction takes place. This could facilitate the development of personalized prediction algorithms tailored to different types of trust dynamics.

In particular, this classification model would help identify individuals who behave differently from a Bayesian decision maker. The decision tree demonstrates high recall ratios for both disbelievers (88\%) and oscillators (100\%). As a result, the model could be used to screen individuals who are more likely to exhibit consistently low trust (disbelievers) or significant trust fluctuations (oscillators). This opens avenues for future research, emphasizing the need for tailored trust calibration methods for disbelievers and a more robust trust prediction model for oscillators.

\section{CONCLUSION}
This study examined the relationships between personal characteristics and trust dynamic types and investigated whether these characteristics could predict the trust dynamics users would exhibit. We conducted a human-subject experiment with 130 participants, where each completed 100 trials of a simulated surveillance task assisted by an imperfect automated threat detector. We gathered data on personal characteristics, trust dynamics, behavior, performance, and post-experimental ratings.

We identified three distinct trust dynamics types (BDM, disbeliever, and oscillator).
Disbelievers showed higher \textit{neuroticism} and lower \textit{performance expectancy} and \textit{PAS-high expectations}, while oscillators rated higher scores in \textit{masculinity}, \textit{positive affect}, \textit{extraversion}, and \textit{intellect}. Using the seven key dimensions, a decision tree model was developed to classify users into these types, highlighting the role of specific personal characteristics as significant predictors of trust dynamics type.

Our study has several limitations that should be acknowledged. First, although the sample size in our lab study is relatively large for the fields of human-computer interaction and human factors, the number of participants classified as disbelievers or oscillators remains limited. A larger sample size would provide clearer insights into the distinct characteristics of these two groups and improve the accuracy of the trust dynamics type prediction model. Second, as one of the first studies investigating the relationship between personal characteristics and trust dynamics, this research took an exploratory approach, allowing us to examine a broad range of personal characteristics. However, we did not strictly control for the variance within each dimension. Future studies should focus on specific dimensions, with more controlled variance, to yield more precise insights. Third, the reliability range used in this study was intentionally set near the threshold where automated/autonomous technologies are perceived as useful. Future research should explore trust dynamics across a wider range of reliability levels. Fourth, the input to the clustering algorithm was handpicked based on prior literature. Future studies could investigate the inclusion of other variables as input. Fifth, the trust prediction model did not differentiate between true negative and true positive, and between false negative and false positive. Future studies could focus on enhancing the trust prediction model by adding more hyperparameters. Sixth, the accuracy of the decision tree model is only 70\%. Future research is needed to investigate the use of other more advanced machine learning models.

\bibliographystyle{IEEEtran}
\bibliography{bibliography}

\end{document}